\documentclass[preprint,12pt,3p]{elsarticle}

\usepackage{matlab-prettifier}
\usepackage{graphicx}
\usepackage{amsmath}
\usepackage{parskip}
\usepackage{hyperref}
\usepackage{adjustbox}
\usepackage{setspace}
\hypersetup{ 
backref=true, 
bookmarks=true, 
unicode=false, 
pdftoolbar=true, 
pdfmenubar=true, 
pdffitwindow=false, 
pdfstartview={FitH}, 
pdftitle={}, 
pdfauthor={}, 
pdfsubject={}, 
pdfkeywords={}{}, 
pdfnewwindow=true, 
colorlinks=true, 
linkcolor=BlueViolet, 
citecolor=maroon, 
urlcolor=blue,
}
\usepackage{comment}
\usepackage{subfig}
\usepackage{caption}
\usepackage{amssymb}
\usepackage{gensymb}

\usepackage{natbib}
\biboptions{comma,round}
\setcitestyle{authoryear}

\begin{document}
\begin{frontmatter}

\title{The neglected contributions of Thomas C. Schelling to the economics of climate change}

\author[label2,label3,label4,label5,label6,label7]{Richard S.J. Tol\corref{cor1}\fnref{label9}}
\address[label2]{Department of Economics, University of Sussex, Falmer, United Kingdom}
\address[label3]{Institute for Environmental Studies, Vrije Universiteit, Amsterdam, The Netherlands}
\address[label4]{Department of Spatial Economics, Vrije Universiteit, Amsterdam, The Netherlands}
\address[label5]{Tinbergen Institute, Amsterdam, The Netherlands}
\address[label6]{CESifo, Munich, Germany}
\address[label7]{Payne Institute for Public Policy, Colorado School of Mines, Golden, CO, United States of America}

\cortext[cor1]{Jubilee Building, BN1 9SL, UK}
\fntext[label9]{I am grateful to Claude for implementing DICE2016 in Python, running multiple versions, and drawing graphs.}

\ead{r.tol@sussex.ac.uk}
\ead[url]{http://www.ae-info.org/ae/Member/Tol\_Richard}

\begin{abstract}
The rich have emitted the bulk of greenhouse gases. The poor suffer the bulk of the impacts. Climate change is a transfer from poor to rich. Climate policy is a transfer from rich to poor. Why, Schelling asked, do people in the Global North care about the descendants of people they do not care about? And, assuming they do, are there no better ways to help them than emission reduction? A prominent economist, Schelling posed his Paradox and Conjecture in a series of papers in the 1980s and 1990s. The economics profession has largely ignored his work, instead focusing on Nordhaus' simpler carbon-as-an-externality framing.

\end{abstract}

\end{frontmatter}

\section{Introduction}
William D. \citet{Nordhaus1975} was the first economist to study climate change. Thomas C. \citep{Schelling1983} was the second. Their contributions are very different.

Schelling added four insights, unique at the time and still not fully understood:
\begin{itemize}
    \item Societies change, often more rapidly than climate \citep{Schelling1990}. Vulnerability to climate change is likely to fall as economies develop \citep{Schelling1992}.
    \item Emissions are high in rich countries, impacts in poor countries. Climate change is a transfer from poor to rich. Climate policy is a transfer from the current rich to the descendants of the current poor \citep{Schelling1995}.
    \item The \emph{Schelling Paradox}:\footnote{I believe \citet{Kolstad2012} coined this term.} Policies on international trade and migration indicate that the current rich do not care much about the current poor, yet advocates of stringent climate policy express great concern about the latter's children and grandchildren, who will probably be better off. These are odd preferences \citep{Schelling1995}.
    \item The \emph{Schelling Conjecture}:\footnote{I believe \citet{Anthoff2012} coined this term. There is another Schelling Conjecture: Domestic constraints provide an advantage in international negotiations \citep{Schelling1960}.} Taking such preferences as given, we should ask how best to help the children of the current poor. Development aid may be more effective in reducing the impacts of climate change \citep{Schelling1995}.
\end{itemize}
Schelling asked penetrating questions that are difficult and uncomfortable. He did not answer these questions\textemdash as he freely admitted. \citet{Nordhaus1992}, on the other hand, asked a simpler question\textemdash What would a global, immortal, benevolent social planner do about climate change?\textemdash and answered that question with a simple and elegant, carefully calibrated model.

Many economists have followed in Nordhaus' footsteps, re-examining and refining his work. Schelling has fewer followers, even though his climate work is arguably more pertinent. The \emph{Schelling Conjecture}, an empirical question, is largely unanswered. The \emph{Schelling Paradox}, an empirical, moral, and theoretical question, is mostly ignored. This may well reflect the economics profession's preference for technical analysis of narrow issues.

Schelling raised two further issues that the economics profession did take up. First, \citet{Schelling1984env} and \citet{Schelling1991MIT} discuss the prospects for international cooperation on greenhouse gas emission reduction\textemdash and conclude these are not good. This was elaborated by \citet{Barrett1990, Barrett1994, Barrett1994b},\footnote{\citet{Barrett2012} pays homage to Schelling.} \citet{Hoel1991, Hoel1994}, \citet{Carraro1992, Carraro1993}, \citet{Chander1992, ChanderTulkens1995}, and \citet{NordhausYang1996}, among many others since \citep[e.g., ][]{Battaglini2016}.

Second, \citet{Schelling1996} discusses geoengineering, noting how much cheaper it is compared to other options to slow climate change and the great difficulties in creating an international regulatory framework. Both points are echoed in later studies \citep[e.g., ][]{Barrett2008, Victor2008, Helm2015, Weitzman2015, Moreno-Cruz2025}. \citet{Pezzoli2023} study the impact of geoengineering on the prospects of international cooperation.

The paper proceeds as follows. In the next three sections, I discuss dynamic vulnerability, the \emph{Schelling Paradox}, and the \emph{Schelling Conjecture}, respectively. Section \ref{sc:conclusion} concludes. The appendix is a collection of quotes.

\section{Dynamic vulnerability}
\label{sc:vulnerability}
Schelling starts by making two points: Poorer countries are more vulnerable to climate change. Economic growth reduces vulnerability.

Schelling's first point is undisputed. In 1983, he wrote matter-of-factly about this. It was uncontroversial in his mind, and I am not aware of anyone who disagrees \citep[e.g., ][]{Yohe2002, Adger2003, Adger2006, Fankhauser2014}. \citet{Fankhauser1995} was the first to quantify the disproportionate impacts on developing economies. Later assessments reach the same qualitative conclusion \citep[e.g., ][]{Tol2021anyas}.

Schelling's second point is almost a corollary: If poverty causes vulnerability then economic growth would reduce the impact of climate change. And indeed, Schelling did not feel the need to elaborate. He made this point in one or a few sentences. See the appendix to reread his words.

Yet, the literature has by and large ignored this. In the first version of the DICE model \citep{Nordhaus1992}, the impact of climate change is a function of climate change and climate change only. This assumption was never changed \citep{Barrage2024}. The recently developed GIVE model makes the same assumption \citep{Rennert2022}. The regional version of DICE assumes that poorer countries suffer proportionally higher damages but do \emph{not} get less vulnerable as they grow richer \citep{NordhausYang1996}.

\citet{Sterner2008} argue that economic growth makes societies \emph{more} vulnerable to climate change, an assumption adopted by \citet{Bijgaart2016, Drupp2021, Bremer2021, Tian2019} and \citet{Wang2024DA}. \citet{Budolfson2017, Budolfson2020} and \citet{Safarzynska2022} explore both increasing and decreasing vulnerability. Only the FUND model consistently follows Schelling \citep{Tol1997, Anthoff2010, Hwang2026}. 

This matters. The social cost of carbon is \$58/tC (\$126/tC) higher (lower) if vulnerability increases (decreases) with economic growth than if it is constant \citep{Tol2026sccmeta}. \citet{Diaz2017} find that the assumed income elasticity of vulnerability to climate change explains most of the difference between the social cost of carbon estimated by three integrated assessment models (DICE, FUND, and PAGE).

Nordhaus knew Schelling's work; they served on committees together and frequented the same workshops. There is no record of disagreement. Nordhaus (personal communication, a long time ago) assumed constant relative vulnerability for a technical reason: The optimization problem is not necessarily convex if economic growth raises emissions but reduces vulnerability: There could be multiple local optima. Constant vulnerability, on the other hand, guarantees the existence of a single, global optimum\textemdash as does \emph{increasing} vulnerability.

\begin{figure}
    \centering
    \caption{Net present welfare plotted against perturbations of the optimal emission control rate for alternative assumptions on the income elasticity of the impact of climate change.}
    \label{fig:dice}
    \includegraphics[width=1.0\linewidth]{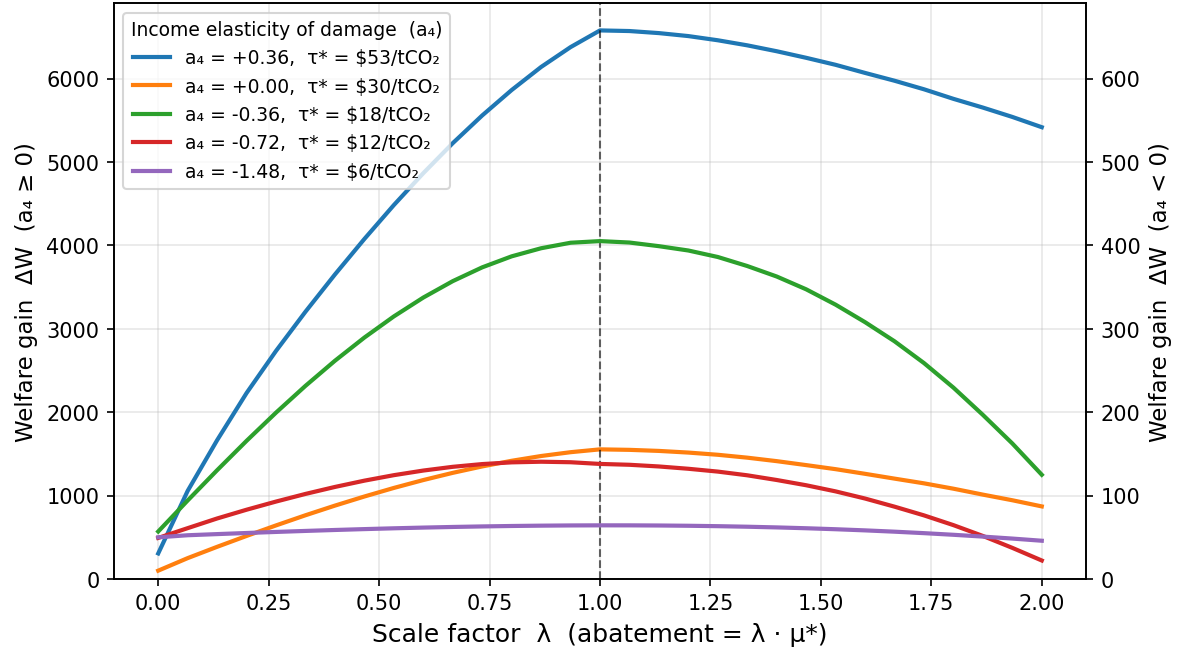}
    \caption*{\footnotesize The model is a Python version of DICE2016 that almost exactly reproduces \citet{Nordhaus2017} for $a_4=0$. The income elasticity $a_4=-0.36$ is taken from \citet{Botzen2021}. The factor $\lambda$ on the horizontal axis scales the optimal emission control rate $\mu$ for all time periods. The vertical axes show the increase in net present welfare relative to the business-as-usual scenario.}
\end{figure}

Figure \ref{fig:dice} shows that DICE2016 \citep{Nordhaus2017} has a unique, global optimum, also if vulnerability falls with economic growth. The reason is that greenhouse gas emission reduction is not sufficiently expensive to affect economic growth such that vulnerability to climate change materially increases. Nordhaus' technical concern is a red herring for his calibration.

Many integrated assessment models adopt Nordhaus' specification of the damage function of climate change, either for the same reason as Nordhaus did or because that is the path of least resistance if you want your paper to focus on some other aspect of the economics of climate change. Furthermore, it may be easier to publish a paper that finds that climate change is worse than previously thought \citep{Tol2025anyas}; dynamic vulnerability \emph{lowers} the social cost of carbon \citep{Diaz2017} and so the chance of getting published.  

\section{The Schelling Paradox}
\label{sc:paradox}
Schelling wondered why some people in the Global North are keener to reduce emissions than to alleviate poverty in the Global South. Why do they care about the children and grandchildren of the people they do not seem to care about? Stingy international aid suggests little concern for the poor. A study of international trade policy and, particularly, the enforcement of restrictions on labour migration could lead to the conclusion that people in the Global North quite dislike the people in the Global South.

There are four reasons, none satisfactory, that could resolve the Schelling Paradox. First, climate change is salient and poverty is not. The former is fresh in our minds, with daily reminders in the media. We are numb to the latter, or prefer not to think about it since it appears to be an intractable problem. The difference in salience may explain the attitude of people\textemdash this hypothesis has not been tested as far as I know\textemdash but it does not justify policy. Policy analysis typically includes the first-best, even if the focus is on deviations from the optimum. Policy advice aims to make policy rational\textemdash rather than give in to irrationality.

The other three reasons are moral. You may argue that the unborn cannot fend for themselves and therefore deserve greater protection than their parents \citep{Bohman2011}. This would justify focusing on intergenerational problems over contemporary ones. However, protecting the child at the mother's expense is not good for the child. Future income levels are based on current incomes. I return to this argument in the discussion of the Schelling Conjecture. \citet{Caney2008} and \citet{Davidson2008} argue that climate change violates the human rights of future generations\textemdash but poverty violates the human rights of their ancestors.

As a third reason for the Schelling Paradox, you may argue that sins of commission\textemdash emitting greenhouse gases\textemdash are worse than sins of omission\textemdash failing to act on poverty. Christian theology has long argued that both are wrong. Criminal law agrees, although it does have gradations, for example, between murder and manslaughter. Following \citet{Coase1960}, you may argue that commission and omission are a matter of perspective: Failing to reduce emissions is a sin of omission, while preventing migrants from crossing borders and sending remittances home is a sin of commission.

Fourth, you may argue that the irreversibility of climate change gives it a special place in any ethical calculus. This too is a stretch. Irreversibility combined with uncertainty and learning qualitatively changes the optimal greenhouse gas emission trajectory \citep{Manne1991, Kolstad1994, Ingham2007}, but this result does \emph{not} hinge on the exceptional status of any group or generation. In other words, it is not an ethical argument, and therefore no response to Schelling. Besides, poverty traps are an important part of underdevelopment. Poverty traps are, by definition, irreversible.

\citet{Wade2009} argue that decisions about future generations are less skewed by selfishness \citep[cf.][]{Buntaine2018} than decisions about the current one. This holds in the very long-term, but less so over the next few generations, which is the time-scale relevant for climate policy \citep{Nesje2024}. Emotional ties are often strong between grandparents and grandchildren, and economic status is passed down the generations.

In short, these attempts to explain away the Schelling Paradox do not stand up to scrutiny.

There is a substantial literature on one ethical dimension of climate change: Time preference\textemdash indeed, \citet{Schelling1995} is most frequently cited in the discounting literature.\footnote{For example, in his commentary on \citet{Schelling1999}, \citet{Rothenberg1999} completely ignores the redistribution emphasized by Schelling.} The literature on risk and, to a lesser extent, ambiguity is sizeable too. There is a smaller literature on inequity aversion, which actually \emph{increases} the weight placed on the future poor relative to the current poor.

The profession's focus on efficiency partly explains economists' reluctance to analyse the Schelling Paradox\textemdash indeed, some maintain that economics is limited to the First Welfare Theorem, the Second Welfare Theorem being the domain of politics and philosophy. Because of this, and again following Nordhaus' lead, climate policy analysis has largely been defined as the study of a single externality and the restoration of Pareto optimality. This requires a definition of the Pareto optimum\textemdash that is, a specification of how to aggregate outcomes over time, over states of the world, and over countries and people\textemdash but refrains from passing judgement on the status quo \citep[cf.][]{Shukla1999}.

A research programme to internalize an externality also does not need to answer the question why this particular externality demands our attention over other problems. The climate economics literature is heavily skewed towards authors from the Global North \citep{Dong2024}. Schelling's Paradox thus remains unstudied. To the best of my knowledge, none of the papers that refer to Schelling's work, see Table \ref{tab:cites}, takes up this challenge.

\section{The Schelling Conjecture}
\label{sc:conjecture}
Schelling \citep[cf.][]{Mendelsohn2012schelling} argued that there are alternatives to greenhouse gas emission abatement to reduce the disproportionate impact of climate change on developing countries. This large vulnerability is partly due to the structure of the economy\textemdash the agricultural sector is more exposed to climate change than the services sector\textemdash and partly due to the capacity to adapt to climate change. Private adaptation depends primarily on access to technology. Public adaptation is almost synonymous with state capacity, including the ability to raise taxes and provide public goods \citep{Yohe2002, Besley2009}. Economic growth would thus reduce vulnerability to climate change, as would development aid targeted at certain public goods\textemdash such hurricane shelters or irrigation infrastructure\textemdash and research into particular technologies\textemdash such as malaria vaccines and weather index insurance.

The various elements of the Schelling Conjecture are active areas of research in economics. Structural change has long been on the agenda, its interaction with climate a more recent addition \citep{Bretschger2011, Nath2025, Wang2025}. Differential vulnerability is well-studied, particularly with regard to natural disasters \citep{Toya2007, Cavallo2011, Klomp2014meta}. Long time-series shine light on the dynamics of vulnerability \citep{Chevet2011, Jongman2015, Barreca2015}. Investment in adaptation is another line of research \citep{Bruin2009, Millner2015, Filatova2025}.

Yet, few papers pull this material together to study the trade-off between investments to reduce emissions and to reduce vulnerability abroad \citep{Tol2005EDE, Tol2007miti, Anthoff2012}. These papers offer a mixed conclusion: A dollar spent on development aid reduces impacts more than a dollar spent on emission reduction in some but certainly not all countries and circumstances. The model used in these three papers is crude. Development aid is an untargeted transfer of income, rather than specific support for health, education, or governance. Directed R\&D is not included.

Two programmes were particularly successful in reducing vulnerability to climate change: Storm shelters and bed nets. The death toll from hurricanes and floods in Bangladesh fell from hundreds of thousands per event to hundreds, primarily because of a system of early warning and shelters \citep{Paul2009, Haque2012}. Across Africa, the annual death toll from malaria fell by half, a smaller relative reduction but a bigger absolute life-saver \citep{Bhatt2015}. Both programmes succeeded in the face of substantial population growth, and were to a substantial extent funded by foreign donors. Hurricanes, floods, and vector-borne diseases are likely to get worse with climate change. Neither programme was primarily about climate change, but vulnerability was reduced in a major way. Not including interventions like these, \citet{Tol2005EDE} probably biased his analysis against the Schelling Conjecture.

At the same time, the shelters in Bangladesh reduced the death toll but not the material damage of floods, which remains a major issue. As noted above, state capacity is a key component of adaptive capacity. It remains lacking in Bangladesh as well as in other countries that are very vulnerable to flooding, such as C\^{o}te d'Ivoire and Nigeria. The Great Green Wall of Africa would have halted desertification but funding either vanished or was misspent \citep{Turner2021, Turner2023}. Successful adaptation requires regime change. \citet{Tol2005EDE} assume steady progress funded by foreign donations, which may be too optimistic\textemdash and a bias in favour of the Schelling Conjecture.

The reasons for the lack of attention to the Schelling Conjecture are the same as those for assuming static vulnerability, see Section \ref{sc:vulnerability}, and ignoring the Schelling Paradox, see Section \ref{sc:paradox}. Investment in adaptation abroad is an additional control variable, complicating the analysis beyond the study of a single externality. You cannot include development aid without confronting the Schelling Paradox.

\section{Discussion and conclusion}
\label{sc:conclusion}
William \citet{Nordhaus1975} was the first economist to write about climate change. Ralph \citet{dArge1979} was the second. His contribution was absorbed by \citet{Nordhaus1991EJ}, whose work has dominated the field ever since. Thomas \citet{Schelling1983} was the third economist to study climate change. His papers are cited, but his central messages ignored.

Schelling's genius was to turn a few simple facts\textemdash emissions are from the past and the North, impacts are in the future and South\textemdash into penetrating questions\textemdash why do we care about the children of the current poor and, if indeed we do, is there no better way to help them? Although these questions were asked decades ago by one of the most prominent economists of the second half of the 20th century, the economics profession has been unable or unwilling to find answers.

This is unfortunate. Important questions should be analysed, not ignored. \citet{Tol2005ESP} argues that mitigation and adaptation occur at different scales and involve different actors. This makes a coherent analysis of the Schelling Conjecture hard\textemdash but not impossible. Bright minds should relish the Schelling Paradox.

Nordhaus's genius was to transform a seemingly intractable problem into a simple externality \citep{Bator1958} with a simple solution: A \citet{Pigou1920} tax. That said, carbon dioxide is the mother of all externalities\textemdash ubiquitous and inequitable in its causes and consequences, global in scope, long-term in nature, its understanding uncertain, ambiguous, and incomplete\textemdash and any estimate of the optimal carbon price has so many degrees of freedom that Nordhaus' simple externality has kept economists busy for decades. Only recently, a student of Nordhaus added an interaction with a second market imperfection \citep{Barrage2020}. Using \citet{Negishi1960} weights, \citet{NordhausYang1996} reinforced the casting of climate change as an \emph{efficiency} problem. Nordhaus' debates, first with William \citet{Cline1992} and later with Nicholas \citet{Stern2006}, focused the \emph{equity} discussion on the choice of the discount rate\textemdash and seems to have led many to believe that Schelling's work was primarily about discounting.

The historical understanding, perhaps explanation, developed above of how the Schelling Paradox and, to a lesser extent, the Schelling Conjecture came to be ignored by the economics profession studying climate policy does not justify these omissions. Indeed, it is about time that climate economists engage with Schelling's work.

\begin{table}[]
    \centering
    \caption{Citations to key climate papers by Thomas C. Schelling}
    \label{tab:cites}
    \begin{tabular}{l r r r} \hline
        Paper & Web of Science & Scopus & Google Scholar \\ \hline
         \citet{Schelling1983} & - & - & -\\
         \citet{Schelling1984env} & 5 & 5 & 19 \\
         \citet{Schelling1984} & 6 & 7 & 132 \\
         \citet{Schelling1990} & 3 & 5 & 28\\
         \citet{Schelling1992} & 228 & - & 772 \\
         \citet{Schelling1995sci} & 1 & 0  & - \\
         \citet{Schelling1995} & 171 & 260 & 558\\
         \citet{Schelling1999} & - & - & - \\
         \citet{Schelling2000} & 22 & 28 & 77 \\ \hline
    \end{tabular}
    \caption*{\footnotesize \citet{Schelling1984} summarizes \citet{Schelling1984env}, which is a reprint of \citet{Schelling1983}. \citet{Schelling1999} and \citet{Schelling2000} are abridged versions of \citet{Schelling1995}.}
\end{table}

\bibliography{master}

\appendix

\newpage \section{Schelling in his own words}
\label{sc:quotes}

\subsection{Vulnerability}
\citet{Schelling1984}:
\begin{quote}
    The most likely possibility [...] is that the impact of climate change on global income and production [...] would not be of alarming magnitude. Particular regions or countries, especially those dependent on agriculture for a large part of their earnings, could be severely affected. The result would look more like a redistribution of global income than a large subtraction from it.
\end{quote}

\citet{Schelling1984env}
\begin{quote}
    If we think of how life has changed in our own country in the last century [...] and suppose that comparably dramatic changes may occur at the same rate in the future [...], it is evident that people will be making multitudinous adaptations in the ways they live and work.
\end{quote}

\citet{Schelling1992}:
\begin{quote}
    It is with the less-developed countries that we have to be most careful about superimposing the climates of the future on the economies and societies of today. As it was in our own country during this century, the trend in developing countries is to be less dependent on agriculture and less vulnerable to climate in transportation and other activities and health. If per capita income growth in the next 40 years compares with the 40 years just past, vulnerability to climate change should diminish, and the resources available for adaptation should be greater.
\end{quote}

\citet{Schelling1995}:
\begin{quote}
    [I]f most of the economic sacrifices in the interest of carbon abatement are borne by the countries that can best afford it, the transfers will tend to be from the well-to-do people of Western Europe, North America, and Japan to the residents of what we now call the 'developing' countries, who should be far better off a century from now than they are now.
\end{quote}

\citet{Schelling1995sci}:
\begin{quote}
    [S]uperimposing [...] future climate changes on today's less-developed societies may lead to an exaggerated assessment of their vulnerabilities after another half-century of economic development.
\end{quote}

\subsection{Paradox}
\citet{Schelling1984}:
\begin{quote}
    Those concerned with the future welfare of Bangladesh [...] have to be more concerned about the floods that will occur during the next 20 or 30 years than the floods that may occur during the 20 or 30 years after the middle of the next century. If the developed countries were prepared to make substantial economic sacrifices now to help provide a more benign climate for Bangladesh a hundred years from now, anyone responsible for Bangladesh would probably prefer to have those economic sacrifices take the form of more immediate economic contributions to the country's standard of living and economic growth.
\end{quote}

\citet{Schelling1990}:
\begin{quote}
    If the more prosperous nations were prepared to help Bangladesh at great expense to themselves, aid now would probably appeal more to Bangladesh than heroic efforts to forestall floods a century hence.
\end{quote}

\citet{Schelling1992}:
\begin{quote}
    A strong argument for trying seriously to slow climate change is that the developing countries are vulnerable and we care.[...] I believe that if the developed countries were prepared to invest [...] in greenhouse-gas abatement, explicitly for the benefit of developing countries 50 years or more from now, the developing countries would clamor to receive the resources immediately in support of their continued development.
\end{quote}

\citet{Schelling1995}:
\begin{quote}
    The optimization models have no provision for redistributing current income. They redistribute only forward in time; contemporary Chinese get nothing from us, but future Chinese we treat as part of the family.
\end{quote}

\citet{Schelling1995}:
\begin{quote}
    To invest resources now in reduced greenhouse emissions is to transfer consumption from ourselves [...] for the benefit of people distant in the future. It is very much like making sacrifices now for people who are distant geographically or distant culturally.
\end{quote}

\citet{Schelling1995}:
\begin{quote}
    It would be strange to forgo a per cent or two of GNP for 50 years for the benefit of Indians, Chinese, Indonesians and others who will be living 50 to 100 years from now\textemdash and probably much better off than today's Indians, Chinese, and Indonesians\textemdash and not a tenth of that amount to increase the consumption of contemporary Indians, Chinese, and Indonesians.
\end{quote}

\citet{Schelling1995}:
\begin{quote}
    [T]oday's undeveloped populations have stronger claims, on the basis of marginal utility, than the populations two or four generations in the future.
\end{quote}

\citet{Schelling1999}:
\begin{quote}
    [T]he beneficiaries [of greenhouse gas abatement] will mainly be the descendants of people in the currently poorer countries.
\end{quote}

\citet{Schelling1999}:
\begin{quote}
    I see little evidence [...] that people want to make significant additional sacrifices to raise living standards among the people who live now in the developing world. It would surprise me if they could get excited about raising living standards in those same parts of the world at a time in the future where those living standards will be [...] substantially elevated over there they are now.
\end{quote}

\subsection{Conjecture}
\citet{Schelling1984env}
\begin{quote}
    There is also a widespread methodological preference for preventive rather than meliorative programs and for dealing with causes rather than symptoms. But it would be wrong to commit ourselves to the principle that if fossil fuels and carbon dioxide are where the problem arises, that must also be where the solution lies.
\end{quote}

\citet{Schelling1990}:
\begin{quote}
    This situation may demand foreign aid to the poorest countries. I would neither expect nor recommend foreign aid directly related to hardships induced by climate change, but rather aid to the poorest.
\end{quote}

\citet{Schelling1992}:
\begin{quote}
    Some environmentalists argue that developing countries should sacrifice some of their hopes for economic development in the interest of slowing the climate change that may prove disastrous. But the advice contains a contradiction. Any disaster to developing countries from climate change will be a disaster to their economic development. What is desired is to optimize development by investing in greenhouse-gas abatement only when that appears, subject to all the uncertainties, to contribute more to their development in the future than the alternative direct investment in development. It is not economic growth versus environment; it is growth with the environment taken into account.
\end{quote}

\citet{Schelling1995}:
\begin{quote}
    [N]o framework for considering the benefits and costs of greenhouse abatement can isolate itself from the opportunity cost: direct investment in the economic improvement of the undeveloped countries.
\end{quote}

\citet{Schelling1995}:
\begin{quote}
    what mix of programs maximizes [...] their [...] utility [...]: greenhouse abatement, direct investment in their economic improvement, or direct subsidies to their consumption?
\end{quote}

\citet{Schelling1999}:
\begin{quote}
    If the people in western Europe, North America, and Japan are to be persuaded to sacrifice some material welfare in the coming decades for beneficiaries in the rest of the world, there is a choice between investing in immediate economic improvement in those countries or investing in improvement that will begin to be felt fifty years from now.

    [R]aising material welfare now [...] meets a more urgent need. [E]conomic development may be the best defense against any possible adverse effects of climate change. We must always consider, when investing in greenhouse gas abatement for the benefit of those future people, the opportunity cost of investing now in more rapid development for the benefit not only of those future people but their equally worthy and more needy ancestors.
\end{quote}

\end{document}